\documentclass[10pt]{article}
\usepackage{fullpage,latexsym,amssymb}
\usepackage[centertags]{amsmath}
\title{Existence and Newtonian limit of nonlinear bound states in
the Einstein-Dirac system}
\author{David Stuart
\\{\it{\small Centre for Mathematical Sciences, Wilberforce Road, 
Cambridge, CB3 OWA,
England}}\\{\it\small{ email:dmas2@cam.ac.uk}}}
\date{} 

\begin{document}
\newcommand{\dirop}{\partial\hspace{-1.25ex}\slash}
\newcommand{\Dirop}{{D}\hspace{-1.55ex}\slash}
\newtheorem{lemma}{Lemma}
\newtheorem{theorem}[lemma]{Theorem}
\newtheorem{corollary}[lemma]{Corollary}
\newtheorem{definition}[lemma]{Definition}
\newtheorem{remark}[lemma]{Remark}
\newtheorem{Notation}[lemma]{Notation}
\newcommand{\proof}{\noindent {\it Proof}\;\;\;}
\newcommand{\qed}{\protect~\protect\hfill $\Box$}

\newcommand{\be}{\begin{equation}}
\newcommand{\ee}{\end{equation}}
\newcommand{\ba}{\begin{eqnarray}}
\newcommand{\ea}{\end{eqnarray}}
\newcommand{\bes}{\[}
\newcommand{\ees}{\]}
\newcommand{\bas}{\begin{eqnarray*}}
\newcommand{\eas}{\end{eqnarray*}}
\newcommand{\hoa}{{H^{1}_{\mbA}}}
\newcommand{\hta}{{H^{2}_{\mbA}}}
\newcommand{\linf}{{L^\infty}}
\newcommand{\tp}{{\tilde P}}
\newcommand{\cf}{{\cal F}}
\newcommand{\ch}{{\cal H}}
\newcommand{\cfnd}{{\cal F}^{n}-{\cal F}^{n-1}}

\newcommand{\spsi}{_{{{\Psi}}}}
\newcommand{\pt}{\frac{\partial}{\partial t}}
\newcommand{\pxk}{\frac{\partial}{\partial {x^k}}}
\renewcommand{\theequation}{\arabic{equation}}
\newcommand{\rgt}{\rightarrow}
\newcommand{\lngrgt}{\longrightarrow}
\newcommand{\intsT}{ \int_{0}^{T}\!\!\int_{\Sigma} }
\newcommand{\dxdt}{\;dx\,dt}
\newcommand{\sublt}{_{L^2}}
\newcommand{\sublf}{_{L^4}}
\newcommand{\naf}{\nabla_\mbA \Phi}
\newcommand{\covt}{(\pt -iA_0) }
\newcommand{\ano}{A^{n}_{0}}
\newcommand{\aNo}{A^{N}_{0}}
\newcommand{\Psino}{\Psi^{n}_{0}}
\newcommand{\PsiNo}{\Psi^{N}_{0}}
\newcommand{\Nmo}{{N\!-\!1}}
\newcommand{\nmo}{{n-1}}
\newcommand{\nmt}{{n-2}}
\newcommand{\Nmt}{{N\!-\!2}}
\newcommand{\gotm}{\frac{\gamma}{2\mu}}
\newcommand{\ootm}{\frac{1}{2\mu}}
\newcommand{\tloc}{T_{loc}}
\newcommand{\tmax}{T_{max}}
\font\msym=msbm10
\def\Real{{\mathop{\hbox{\msym \char '122}}}}
\font\smallmsym=msbm7
\def\smr{{\mathop{\hbox{\smallmsym \char '122}}}}
\def\Complex{{\mathop{\hbox{\msym\char'103}}}}
\newcommand{\wkarr}{\; \rightharpoonup \;}
\def\Weak{\,\,\relbar\joinrel\rightharpoonup\,\,}
\newcommand{\To}{\longrightarrow}
\newcommand{\rp}{\hbox{Re\,}}
\newcommand{\pa}{\partial_A}
\newcommand{\pbfa}{\partial_{\mathbf A}}
\newcommand{\pao}{\partial_{A_1}}
\newcommand{\pat}{\partial_{A_2}}
\newcommand{\dbar}{\bar{\partial}}
\newcommand{\barpa}{\bar{\partial}_{A}}
\newcommand{\barpbfa}{\bar{\partial}_{\mathbf A}}
\newcommand{\barpaphi}{\bar{\partial}_{A}\Phi}
\newcommand{\barpbfaphi}{\bar{\partial}_{\mathbf A}\Phi}
\newcommand{\myqed}{\hfill $\Box$}
\newcommand{\cd}{{\cal D}}
\newcommand{\dt}{\hbox{det}\,}
\newcommand{\sma}{_{{ A}}}
\newcommand{\bfpi}{{\mbox{\boldmath$\pi$}}}
\newcommand{\ce}{{\cal E}}
\newcommand{\ulh}{{\underline h}}
\newcommand{\ulg}{{\underline g}}
\newcommand{\ulX}{{\underline {\bf X}}}
\newcommand\la{\label}
\newcommand{\lamo}{\stackrel{\circ}{\lambda}}
\newcommand{\bfjo}{\underline{{\bf J}}}
\newcommand{\Vflato}{V^\flat_0}
\newcommand{\cm}{{\cal M}}
\newcommand{\dist}{{\mbox{dist}}}
\newcommand{\cs}{{\cal S}}
\newcommand{\mcF}{{\mathcal F}}
\newcommand{\mcn}{{\mathcal  V}}
\newcommand{\mce}{{\mathcal  E}}
\newcommand{\mcb}{{\mathcal B}}
\newcommand{\mca}{{\mathcal A}}
\newcommand{\mcdpsi}{{\mathcal D}_{\psi}}
\newcommand{\mcd}{{\mathcal D}}
\newcommand{\mcl}{{\mathcal L}}
\newcommand{\mclaphi}{{\mathcal L}_{(\mbA,\Phi)}}
\newcommand{\mcdadjp}{\mcdadj_{\psi}}
\newcommand{\mcdadj}{{\mathcal D}^{\ast}}
\newcommand{\zl}{{Z_\Lambda}}
\newcommand{\thl}{{\Theta_\Lambda}}
\newcommand{\ca}{{\cal A}}
\newcommand{\cb}{{\cal B}}
\newcommand{\cg}{{\cal G}}
\newcommand{\cu}{{\cal U}}
\newcommand{\co}{{\cal O}}
\newcommand{\smA}{\small A}
\newcommand{\ttheta}{\tilde\theta}
\newcommand{\tn}{{\tilde\|}}
\newcommand{\rbar}{\overline{r}}
\newcommand{\oeps}{\overline{\varepsilon}}
\newcommand{\cgl}{\hbox{Lie\,}{\cal G}}
\newcommand{\Ker}{\hbox{Ker\,}}
\newcommand{\const}{\hbox{const.\,}}
\newcommand{\Sym}{\hbox{Sym\,}}
\newcommand{\tr}{\hbox{tr\,}}
\newcommand{\grad}{\hbox{grad\,}}
\newcommand{\ttd}{{\tt d}}
\newcommand{\ttdel}{{\tt \delta}}
\newcommand{\ns}{\nabla_*}
\newcommand{\csl}{{\cal SL}}
\newcommand{\kr}{\hbox{Ker}}
\newcommand{\beq}{\begin{equation}}
\newcommand{\eeq}{\end{equation}}
\newcommand{\pr}{\hbox{proj\,}}
\newcommand{\proj}{{\mathbb P}}
\newcommand{\tproj}{\tilde{\mathbb P}}
\newcommand{\projq}{{\mathbb Q}}
\newcommand{\tprojq}{\tilde{\mathbb Q}}
\newcommand{\oN}{\overline N}
\newcommand{\cN}{\cal N}
\newcommand{\cmet}{{{\hbox{${\mathcal Met}$}}}}
\newcommand{\met}{{\hbox{Met}}}
\newcommand{\bfga}{{\mbox{\boldmath$\overline\gamma$}}}
\newcommand{\bfOmega}{\mbox{\boldmath$\Omega$}}
\newcommand{\bfTh}{\mbox{\boldmath$\Theta$}}
\newcommand{\bfmw}{{\bf m}_{\mbox{\boldmath$\smo$}}}
\newcommand{\bfmu}{{\mbox{\boldmath$\mu$}}}
\newcommand{\bfulX}{{\mbox{\boldmath${X}$}}}
\newcommand{\bfultX}{\tilde{\mbox{\boldmath${X}$}}}
\newcommand{\bfmuw}{{\mbox{\boldmath$\mu$}}_{\mbox{\boldmath$\smo$}}}
\newcommand{\dmug}{d\mu_g}
\newcommand{\ltwo}{{L^{2}}}
\newcommand{\lfour}{{L^{4}}}
\newcommand{\hone}{H^{1}}
\newcommand{\honea}{H^{1}_{\smA}}
\newcommand{\er}{e^{-2\rho}}
\newcommand{\onetwo}{\frac{1}{2}}
\newcommand{\lra}{\longrightarrow}
\newcommand{\dv}{\hbox{div\,}}
\newcommand{\mbA}{\mathbf A}
\def\smo{{\mbox{\tiny$\omega$}}}
\protect\renewcommand{\theequation}{\thesection.\arabic{equation}}

\font\msym=msbm10
\def\Real{{\mathop{\hbox{\msym \char '122}}}}
\def\R{\Real}
\def\Z{\mathbb Z}
\def\K{\mathbb K}
\def\J{\mathbb J}
\def\L{\mathbb L}
\def\D{\mathbb D}
\def\Mink{{\mathop{\hbox{\msym \char '115}}}}
\def\Integers{{\mathop{\hbox{\msym \char '132}}}}
\def\Complex{{\mathop{\hbox{\msym\char'103}}}}
\def\C{\Complex}
\font\smallmsym=msbm7
\def\smr{{\mathop{\hbox{\smallmsym \char '122}}}}

\maketitle
\thispagestyle{empty}
\vspace{-0.3in}
\begin{abstract}
An analysis  is given of particlelike nonlinear bound states in
the Newtonian limit of the coupled Einstein-Dirac system
introduced by Finster, Smoller and Yau. A proof is given of existence of
these bound states in the almost Newtonianian regime, 
and it is proved that they may be
approximated by the energy minimizing solution of the Newton-Schr\"odinger
system obtained by Lieb.
\end{abstract}

\section{{\large Introduction}}
\label{secint}

Since the gravitational interaction is attractive, there is the possibility
that theories involving fields coupled to gravity will possess bound states
held together by this force. At the level of Newtonian gravity an example
is provided by the Newton-Schro\"odinger system, in which there are
bound states which can be obtained by minimizing a nonlocal energy functional
(\cite{lieb}); see also more recent discussions in \cite{hmt03,p96,cs}. 
In general relativity, it was first established that this phenomenon occurs for
the Yang-Mills-Einstein equations (see \cite{bmck,sw} for the original work, 
and \cite{bfm} for some more
recent developments and further references) and then for various
other systems involving Dirac fields (\cite{fsy99,fsy99b}). In this article
we discuss the bound states in the Einstein-Dirac system which were
studied numerically in \cite{fsy99}, and prove that in the almost
Newtonian regime
they may be rigorously approximated by the bound states in the 
Newton-Schr\"odinger system, and as a consequence give an existence proof.

The starting point is the action functional
\beq\la{act}
S=\frac{c^3}{8\pi G}\int R\,d\mu_g+\sum\limits_{A=1}^2
\hbar\int\overline{\Psi}_A\bigl(\Dirop  -\frac{mc}{\hbar}  \bigr)\Psi_A\,d\mu_g
\eeq
describing the interaction of two Dirac spinor fields
$\Psi_1$ and $\Psi_2$ with a gravitational metric $g$, whose 
scalar curvature is $R$ and whose volume element is $d\mu_g$. 
We use $\Dirop$ for the Dirac operator derived from $g$, with 
associated $\gamma$ matrices, as explained in 
\cite[Section II]{fsy99}.
The Euler-Lagrange equations are
\beq\la{eom}
R_{ab}-\frac{1}{2}R g_{ab}=\frac{8\pi G}{c^4}T_{ab},
\qquad(\Dirop-\frac{mc}{\hbar})\Psi_A=0
\eeq
where $R_{ab}$ is the Ricci curvature and 
$T_{ab}=\frac{\hbar c}{2}\sum\rp\bigl[\overline{\Psi}_A
(i\gamma_a\partial_b+i\gamma_b\partial_a)\Psi_A\bigr]$
is the energy-momentum tensor; space-time indices
$a,b$ take values in $\{0,1,2,3\}$.
The system \eqref{eom} is precisely that
studied in \cite{fsy99}, 
except that the dimensional constants
$G,\hbar,c$ have been reinstated.  It is always to be 
understood that $\hbar,m,G$ are fixed positive
numbers, while the speed of light 
$c$ is a positive number taking large values, i.e.
we will study the Newtonian (or non-relativistic) 
limit in which $c\to +\infty$. 
The reason there are two Dirac spinor fields in the model is that this allows
for spherical symmetry: indeed,
in the article \cite{fsy99} a spherically
symmetric ansatz was introduced, and the corresponding system of 
ordinary 
differential equations (ODEs) for static spherically symmetric 
solutions was derived.
Numerical evidence was presented for the existence
of solutions to this system of ODEs, which give particlelike,
or solitonic, solutions
of \eqref{eom} corresponding to nonlinear bound states of 
two fermions; 
also the linear stability of these solutions was investigated.
In this note we will show that in the Newtonian
limit $c\to+\infty$ these solutions of 
the system \eqref{eom} can be 
approximated by solutions of the Newton-Schr\"odinger 
system
\beq\la{ns}
i\hbar\frac{\partial\psi}{\partial t}=-\frac{\hbar^2}{2m}\Delta\psi+u\psi,
\qquad
-\Delta u=-\kappa |\psi|^2
\eeq
for $\kappa=8\pi Gm$. In fact the rigorous analysis of the 
limit we provide yields a proof of existence for the bound state
solutions of \eqref{eom} introduced in \cite{fsy99}, 
which does not seem to have appeared
previously in the literature. The type of solution of \eqref{ns}
is one in which the time dependence  is uniform phase rotation
at frequency $\frac{\eta}{\hbar}$: explicitly
$\psi=e^{-i\frac{\eta }{\hbar}t}\varphi(x)$, and $\varphi$ solves
\beq\la{nss}
\eta\varphi=-\frac{\hbar^2}{2m}\Delta\varphi+m u\varphi,
\qquad
-\Delta u=-8\pi Gm |\varphi|^2.
\eeq
The existence of such nonlinear bound state solutions 
to the Newton-Schr\"odinger system 
was proved in \cite{lieb} by variational methods.
Pseudo-relativistic generalizations of Lieb's solutions
have been given in \cite{lenz08}, and we will make use of a result
from this article on the non-degeneracy of the linearization of \eqref{nss}:
see lemma \ref{lll} and the subsequent remark.

\subsection{\normalsize Spherical symmetry} Using the ansatz 
for bound states with spherical symmetry
introduced in \cite{fsy99}, we search for solutions with metric
\beq
g=c^2e^{2\nu}dt^2-e^{2\lambda}dr^2-r^2 d\Omega^2
\label{anz1}\eeq
and spinor fields of the form
\beq
\Psi_1=e^{\nu/2}e^{-i\omega t}\left(
\begin{array}{c}
\Phi_1 e_1
\\
i\Phi_2\sigma^r e_1
\end{array}
\right),\qquad
\Psi_2=e^{\nu/2}e^{-i\omega t}\left(
\begin{array}{c}
\Phi_1 e_2\\
i\Phi_2\sigma^r e_2
\end{array}
\right)  
\label{anz2}\eeq
where
$$
e_1=\left(\begin{array}{c}
1\\0
\end{array}\right)\quad
e_2=\left(\begin{array}{c}
0\\1
\end{array}\right)\quad\hbox{and}\quad\sigma^r=\frac{1}{r}
\sum_{i=1}^3x^i\sigma^i
$$
where $\sigma^i$ are the Pauli matrices.
In \eqref{anz2} $\nu,\lambda,\Phi_1,\Phi_2$ all depend on 
$r$ only, and
$$
\omega\hbar=mc^2+\eta
$$
with $\eta$ as in \eqref{nss}. 
\begin{remark}
The angular dependence in \eqref{anz2} is taken over directly from
that displayed by the ground state Dirac wave functions for the relativistic
hydrogen atom. Recall from \cite{sak67,tha92} that in problems involving
the Dirac equation with spherical symmetry there are three commuting
operators: $J^2, J_3$ and $K$ (respectively the total angular momentum 
squared, the third component of the total angular momentum, and the 
spin-orbit coupling operator defined in \cite[Equation 3.275]{sak67}) 
which also commute with the Hamiltonian. The
simultaneous eigenspaces of these three operators (corresponding to
eigenvalues conventionally written $j(j+1)\hbar^2,j_3\hbar,\kappa\hbar$) are
two dimensional and give the decomposition into irreducible subspaces. 
The possible values
of $j$ are in the set $\{\frac{1}{2},\frac{3}{2},\dots\}$, those of
$j_3$ are in the set $\{-j,\dots,+j\}$ and $\kappa=\pm (j+\frac{1}{2})$.
The ground state corresponds to the choice $j=\frac{1}{2}$, and the
wave functions $\Psi_1$ (resp. $\Psi_2$) 
correspond to $j_3=\frac{1}{2}$ (resp. $j_3=-\frac{1}{2}$) and $\kappa=-1$.
The value of $\kappa$ is not important, but what is crucial is that
the states $\Psi_1,\Psi_2$ have opposite values of $j_3$, which ensures
that the energy momentum tensor $T_{ab}$ is consistent with the 
spherically symmetric metric in \eqref{anz1}: see \cite[Setion IV]{fsy99}.
\end{remark}

Substitution of \eqref{anz1}-\eqref{anz2} into \eqref{eom} leads,
as in \cite[Section IV]{fsy99}, 
to the following system of equations:
\begin{align}
&\bigl(\frac{\omega e^{-\nu}}{c}-\frac{mc}{\hbar}\bigr)\Phi_1
-e^{-\lambda}\frac{\partial\Phi_2}{\partial r}-
\frac{(e^{-\lambda}+1)}{r}\Phi_2=0,\label{d1}\\
&\bigl(\frac{\omega e^{-\nu}}{c}+\frac{mc}{\hbar}\bigr)\Phi_2
+e^{-\lambda}\frac{\partial\Phi_1}{\partial r}+
\frac{(e^{-\lambda}-1)}{r}\Phi_1=0\label{d2}
\end{align}
to be solved coupled to
\begin{align}
e^{-2\lambda}(2r\lambda'-1)+1&=8\pi G\frac{r^2}{c^2}\rho,\label{e1}\\
e^{-2\lambda}(2r\nu'+1)-1&=8\pi G\frac{r^2}{c^4}p,\label{e2}
\end{align}
where $\rho=\rho(\nu,\Phi_1,\Phi_2,\eta,c)$ 
and $p=p(\nu,\Phi_1,\Phi_2,\eta,c)$
are given by
\begin{align}
\rho&=2(m+c^{-2}\eta)e^{-2\nu}(\Phi_1^2+\Phi_2^2)
\\
p&=-2(m+c^{-2}\eta)e^{-2\nu}(\Phi_1^2+\Phi_2^2)\\
&\qquad+4\frac{\hbar c}{r}e^{-\nu}\Phi_1\Phi_2
+2mc^2 e^{-\nu}(\Phi_1^2-\Phi_2^2)\\
&=-2\eta e^{-2\nu}(\Phi_1^2+\Phi_2^2)-2mc^2e^{-\nu}\Phi_1^2(e^{-\nu}-1)\\
&\qquad-2mc^2e^{-\nu}\Phi_2^2+4\frac{\hbar c}{r}\Phi_1\Phi_2.
\end{align}

\subsection{ \normalsize Newtonian limit} We will 
solve the system
\eqref{d1}-\eqref{e2} in
the almost Newtonian regime. 
To be precise, we treat $\epsilon=c^{-1}$ 
as a small positive parameter, and show that the 
system \eqref{d1}-\eqref{e2} can be
regarded as a perturbation of the 
Newton-Schr\"odinger system \eqref{ns}: 
\begin{quote}
{ \em The spherically symmetric 
Einstein-Dirac system \eqref{d1}-\eqref{e2} admits 
nonlinear bound state solutions 
$(\lambda^\epsilon,\nu^\epsilon,\Phi_1^\epsilon,
\Phi_2^\epsilon)$
for small positive $\epsilon$, 
which can be approximated (in a strong weighted norm)
by the bound state solution $\varphi$
of the Newton-Schr\"odinger system which
minimizes the energy \eqref{nlocen}: in particular, 
$(\Phi^\epsilon_1,\Phi^\epsilon_2)$ converges
uniformly to $(\varphi,0)$ as $\epsilon\to 0$.}
\end{quote}
To formulate more precisely and prove this we carry out
a rescaling of the dependent variables, 
and also make an adjustment 
to facilitate the handling of the ADM mass. The precise statement
appears in theorem \ref{main}, and asserts the validity of the
Newtonian approximation in norms stronger than $C^1$ with
exponential weights in the (massive) 
Dirac fields $\Phi_1,\Phi_2$, and polynomial weights for the
(massless) metric components $\lambda,\nu$: see \eqref{ps2}.
The bound state of the Newton-Schr\"odinger system determines
space and time scales, and the requirement that $\epsilon$ be
small (or, equivalently, that the speed of light $c$ be large)
should be regarded as being relative to the scale 
so determined.

\subsection{\normalsize Auxiliary conditions}
We will work with 
boundary conditions corresponding to an asymptotically Minkowskian
metric, i.e. $$\lim_{r\to +\infty}e^{2\lambda(r)}=1
=\lim_{r\to +\infty}e^{2\nu(r)},$$
and also require finite ADM mass, so that
$$
\lim_{r\to +\infty}\frac{r}{2}(e^{2\lambda(r)}-1)
=\epsilon^2 l
$$ 
exists and is finite (for each positive $\epsilon=c^{-1}$; the
scaling factor $\epsilon^2$ is introduced for later convenience
in analyzing the limit $\epsilon\to 0$). 

As $r\to 0$ the solutions will satisfy $e^{2\lambda}=1+O(r^2)$
and $\Phi_2=O(r)$. The normalization condition 
$
\|\Psi_A\|_{L^2(\{t=constant\})}=1,
$
which ensures that the total probability equals one,
can always be imposed on solutions by rescaling,
see \cite[Section VI]{fsy99}. We will therefore not
impose it throughout this paper, since a simple rescaling
ensures that the solutions we obtain in theorem
\ref{main} do satisfy it.

\subsection{ \normalsize Rescaled variables} We now introduce new 
variables $Q,N,\psi_2$ 
(in place, respectively, of $\lambda,\nu,\Phi_2$) 
which take into account both the auxiliary conditions 
and the expected behaviour in the Newtonian limit:
\beq
\epsilon^{-2}(1-e^{-2\lambda})={2l}f_0(r)+Q\qquad
\epsilon^{-2}\nu=N,\label{nv1}
\eeq
with $f_0(r)=r^2/(1+r)^3$, and
\beq
\Phi_2=\epsilon\psi_2.
\label{nv2}
\eeq
\begin{remark}\label{r2}
The variable $l$ determining the ADM mass is determined dynamically,
as will become clear in the implicit function theorem set-up following
\eqref{dy}. It turns out to be convenient for the analysis,
however, to separate it off by introducing the function $f_0$
and requiring $Q=O(r^{-2})$ as $r\to+\infty$.
Other choices of $f_0$ with the same asymptotic behaviour 
would be possible: the choice made here is just a simple function
having the right asymptotics as $r\to\infty$, and vanishing
to $O(r^2)$ as $r\to 0$ as is natural, see remark \ref{vq}.
Notice that $l$ is only uniquely determined
by the requirement $Q=O(r^{-2})$ as $r\to+\infty$, and this condition
is built into the function spaces used in the implicit function theorem.
\end{remark}

\subsection{ \normalsize Formal consideration of the Newtonian limit} 
Equations \eqref{e1}-\eqref{e2} become
\begin{align}
(rQ)'&+2l (rf_0)'-8\pi Gr^2\rho_\epsilon=0,\label{er1}\\
2rN'&-\frac{2l f_0+Q}{1-\epsilon^2 (2l f_0+Q)}-
\frac{8\pi G\epsilon^2 r^2 p_\epsilon}
{{1-\epsilon^2 (2l f_0+Q)}}=0,\la{er2}
\end{align}
with
$\rho_\epsilon(N,\Phi_1,\psi_2,\eta)=
\rho(\epsilon^2 N,\Phi_1,\epsilon\psi_2,\eta,\epsilon^{-1})$
and
$p_\epsilon(N,\Phi_1,\psi_2,\eta)=
p(\epsilon^2 N,\Phi_1,\epsilon\psi_2,\eta,\epsilon^{-1})$,
while \eqref{d1}-\eqref{d2} become
\begin{align}
&\bigl(\frac{\eta}{\hbar}-\frac{mN}{\hbar}\bigr)\Phi_1
-\frac{\partial\psi_2}{\partial r}-
\frac{2}{r}\psi_2-F_1=0,\label{dr1}\\
&\frac{2m}{\hbar}\psi_2
+\frac{\partial\Phi_1}{\partial r}
-F_2=0,\label{dr2}
\end{align}
where
\begin{align}
F_1&=\frac{m}{\hbar}\frac{(1-\epsilon^2 N-e^{-\epsilon ^2 N})}{\epsilon^2}\Phi_1
\la{f1}\\
&\quad
+\frac{\eta}{\hbar}(1-e^{-\epsilon^2 N})\Phi_1 
-\epsilon^2 \bigl(Q+2lf_0\bigr)
\bigl(\frac{\partial\psi_2}{\partial r}+\frac{1}{r}\psi_2
\bigr)
\notag
\end{align}
and
\begin{align}
F_2=\epsilon^2 (Q+2lf_0)\bigl(\frac{\partial\Phi_1}{\partial r}
+\frac{1}{r}\Phi_1\bigr)-\epsilon^2\frac{\eta}{\hbar} e^{-\epsilon^2 N}\psi_2
+\frac{m}{\hbar}(1-e^{-\epsilon^2 N})\psi_2.
\la{f2}
\end{align}

Note that in the limit $\epsilon
\to 0$, the inhomogeneous terms
$F_1$ and $F_2$ are {\em formally} $O(\epsilon^2)$, while
\begin{align}\la{reps}
\begin{split}
\rho_\epsilon&=2(m+\epsilon^{2}\eta)e^{-2\epsilon^2 N}(\Phi_1^2
+\epsilon^2\psi_2^2)\\
&=2m|\Phi_1|^2+O(\epsilon^2)\\
p_\epsilon&=
-2\eta e^{-2\epsilon^2 N}(\Phi_1^2+\epsilon^2\psi_2^2)
-2m e^{-\epsilon^2 N}\Phi_1^2\frac{(e^{-\epsilon^2 N}-1)}
{\epsilon^{2}}
-2me^{-\epsilon^2 N}\psi_2^2+4\frac{\hbar }{r}
\Phi_1\psi_2\\
&=O(1),
\end{split}
\end{align}
so that \eqref{er1}-\eqref{er2} imply that
$N$ becomes the Newtonian potential 
(i.e. the function $u$ in \eqref{nss}), and (eliminating $\psi_2$ 
between \eqref{dr1}-\eqref{dr2})
the function $\Phi_1$ approaches $\varphi$, 
a solution of the Newton-Schr\"odinger system \eqref{nss}.
So far this is completely formal, 
but it will be made into a rigorous statement in theorem 
\ref{main}
in section \ref{nl}. 
In order to achieve this it is
necessary to recall some
facts about \eqref{nss}.

\section{\large The Newton-Schr\"odinger system}
\la{nes}
We briefly summarize the approach to \eqref{nss} adopted in the
article \cite{lieb} so as to provide sufficient
details for our purposes.
\begin{theorem}[\cite{lieb}]\la{lt}
The nonlocal energy 
associated to \eqref{nss}
\beq\la{nlocen}
\frac{\hbar^2}{2m}\int |\nabla\varphi(x)|^2dx-m^2G\int\int
\frac{|\varphi(x)|^2|\varphi(y)|^2}{|x-y|}dxdy
\eeq
admits a finite lower bound 
subject to the constraint of having $\int|\varphi(x)|^2 dx=1$ 
fixed, and this lower bound is attained on a
function which is unique up to translation. Further
this minimizer is positive, spherically symmetric and
a monotone non-increasing function of the radial coordinate
satisfying $|\varphi(r)|\leq c_1 e^{-c_2 r}$ for some
positive numbers $c_1,c_2$.
\end{theorem}
We summarize the points of the proof from \cite{lieb} which
will be needed below.
\begin{enumerate}
\item
The existence of a spherically symmetric minimizer of the 
nonlocal energy \eqref{nlocen}
is proved by means of the Riesz rearrangement inequality, and
a strict version of this inequality implies that any minimizer
is spherically symmetric.
The corresponding Euler-Lagrange equation is
\beq
\la{ele}
-\frac{\hbar^2}{2m}\Delta\varphi(x)-2m^2G\int
\frac{|\varphi(y)|^2}{|x-y|}dy\,\varphi(x)=\eta\varphi(x)
\eeq
where $\eta<0$ is the Lagrange multiplier.

\item The relation between \eqref{nss} and \eqref{ele}
follows quickly using the condition
$\lim_{|x|\to +\infty}|u(x)|=0$ and the formula for the 
solution of Poisson's equation $-\Delta u=f$ on $\R^3$, namely:
\beq\la{fs}
(-\Delta^{-1}f)(x)=\int \frac{f(y)}{4\pi|x-y|}dy.
\eeq 
\item
For the case when $f(x)=\kappa\rho(r)$,
where $\kappa>0$ is a constant and $\rho$
is a function
of the radial coordinate $r=|x|$ only, a result of Newton
(\cite{ll}) implies that \eqref{fs} can be rewritten:
\begin{align}
(-\Delta^{-1}\kappa\rho)(r)&=
-\kappa\int_0^r \Bigl(\frac{1}{s}-\frac{1}{r}\Bigr)
\rho(s) s^2ds
+\kappa\int_0^\infty \rho(s) sds\notag\\
&=-\kappa\int_0^r \Bigl(\frac{1}{s}-\frac{1}{r}\Bigr)\rho(s) 
s^2ds
+u(0).\notag
\end{align}
Define $K(r,s)\equiv 8\pi s^2(\frac{1}{s}-\frac{1}{r})$; this kernel 
is non-negative for
$0\leq s\leq r$. It is shown in \cite{lieb} that any energy minimizing solution
to \eqref{nss} is radially symmetric, and so solves 
the equation
\beq\la{lf}
E\varphi=-\frac{\hbar^2}{2m}\biggl(\frac{d^2}{dr^2}
+\frac{2}{r}\frac{d}{dr}\biggr)\varphi+m^2 G\biggl(\int_0^r
K(r,s)|\varphi(s)|^2ds\biggr)\varphi
\eeq
where $E=(\eta-m u(0))$. 
\item
It is proved in \cite{lieb} that 
all positive solutions of \eqref{lf} can be obtained 
by a scaling of the unique positive solution of
\beq\la{can}
-\biggl(\frac{d^2}{dr^2}
+\frac{2}{r}\frac{d}{dr}\biggr)\phi+\biggl(\int_0^r
K(r,s)|\phi(s)|^2ds\biggr)\phi=\phi.
\eeq
(The proof of uniqueness of positive solutions of \eqref{can} 
is a crucial part of the article \cite{lieb}).
Conversely, as long as $E>0$, there is a positive solution of
\eqref{lf}, which can be obtained by scaling the 
unique positive solution of \eqref{can}.
\end{enumerate}

\begin{lemma}[\cite{lenz08}]
\la{lll}
Let  $L$ be the linear operator obtained by 
linearizing \eqref{ele}:
\beq\notag
L\chi(x)=\biggl(-\frac{\hbar^2}{2m}\Delta-\eta\biggr)\chi(x)
-2m^2G\int
\frac{|\varphi(y)|^2}{|x-y|}dy\,\chi(x)
-4m^2G\int
\frac{\langle\varphi(y),\chi(y)\rangle}{|x-y|}dy\,\varphi(x).
\eeq
Then $L$ is a linear homeomorphism from $H^2_{rad}$ to $L^2_{rad}$
(where these are the radial parts of the usual Sobolev and 
Lebesgue spaces).
\end{lemma}
\proof
Recall that $\eta<0$. 
It is proved in \cite{lenz08} that the operator $L$ is self-adjoint
on $L^2_{rad}$ with domain $H^2_{rad}$ and trivial kernel
and a single negative eigenvalue with one dimensional 
eigenspace. (Going out of the radial sector, $L$ has a three
dimensional kernel associated with translation invariance,
but this does not concern us in this article.)
From this it follows that there
exists $c>0$ such that $\|L\chi\|_{L^2}\geq c\|\chi\|_{L^2}$,
and that $L$ maps $H^2_{rad}$ continuously
onto $L^2_{rad}$. Furthermore the standard elliptic estimate 
gives $\|L^{-1}f\|_{H^2}\leq C\|f\|_{L^2}$ so 
that $L$ is a linear 
homeomorphism $H^2_{rad}\to L^2_{rad}$.
\myqed
\begin{remark}
Curiously in the present general relativistic problem it is possible to use
diffeomorphism invariance to circumvent the need for this lemma. Briefly,
by rescaling time, it is possible to linearize the system with the value of
the gravitational potential fixed at the origin $r=0$ rather than at
infinity; for this linearization the corresponding nondegeneracy result
is rather easy to prove. This was the approach taken in the first version 
of this article,
and it has the advantage of not needing the more careful analysis of $L$
given in \cite{lenz08}. On the other hand it has the disadvantage of giving
solutions which are not asymptotically Minkowskian (without further co-ordinate
changes), and this makes the physical picture less clear in various ways. 
Therefore it seems
preferable to make use of the results of \cite{lenz08} and obtain solutions
which are immediately 
asymptotically Minkowskian.
\end{remark}
\section{\large Theorem on the Newtonian limit}
\la{nl}
\subsection{ \normalsize The analytic set-up} We now introduce the spaces of functions with
which we will be working. 
Let $BC^\delta=BC^\delta([0,\infty))$ be the space of bounded
continuous functions on the half line $0\leq r<\infty$
with $\|f\|_{BC^\delta}=\sup (1+r)^{\delta}|f(r)|<\infty$; 
notice that elements 
of $BC^\delta$ satisfy $f=O(r^{-\delta})$ as $r\to \infty$. 
Let $BC^\delta_{\delta'}$ be the subspace of $BC^\delta$ consisting
of functions $f$ vanishing to order $\delta'>0$ at the origin, 
in the sense that
$$
\|f\|_{BC^\delta_{\delta'}}
=\sup r^{-\delta'}|f(r)|+\sup (1+r)^{\delta}|f(r)|<\infty.
$$
Further, let
$BC^{1,\delta}=BC^{1,\delta}([0,\infty))$ be the subspace of 
$BC^\delta$
consisting of functions $f$ which are also
continuously differentiable with $f'\in BC^{1+\delta}$,
with the norm 
$$\|f\|_{BC^{1,\delta}}=\sup (1+r)^{\delta}|f(r)|
+\sup (1+r)^{1+\delta}|f'(r)|<\infty,$$ and, for $\delta'\geq 1$
let $BC^{1,\delta}_{\delta'}\subset BC^{1,\delta}$ be the subset of
those vanishing at the origin to order $\delta'$,
with norm 
$\|f\|_{BC^{1,\delta}_{\delta'}}=\|f\|_{BC^{1,\delta}}
+\|f\|_{BC^\delta_{\delta'}}
+\|f'\|_{BC^{\delta+1}_{\delta'-1}}<\infty$.
Let $H^s_{rad}$ and
$L^2_{rad}$ be the radial parts of the usual 
Sobolev and Lebesgue spaces. 
We will use without comment the following inequalities: there
exist positive constants $C_1,C_2,C_3,C_4,R_*$ such that
\begin{align}
\sup_{x\in\smr^3}|f(x)|&\leq C_1\|f\|_{H^2(\smr^3)}\la{s1}\\
\|f\|_{L^6(\smr^3)}&\leq C_2\|f\|_{H^1(\smr^3)}\la{s2}\\
\|f/|x|\|_{L^2(\smr^3)}&\leq C_3\|f\|_{H^1(\smr^3)}\la{s3}\\
\sup_{x\in\smr^3:|x|\geq R_*}|xf(x)|&\leq
C_4\|f\|_{H^1(\smr^3)}\,\qquad
\hbox{for {\em radial} functions $f$}\la{s4}
\end{align}
(see \cite{strauss}, or \cite{bl}), and also the fact that
$H^2_{rad}(\R^3)$ functions
are continuous and $O(r^{-1})$ as $r\to+\infty$.
The inequality \eqref{s3} is
Hardy's inequality. The $\;_{rad}$ suffix will be ommitted
when writing a norm.

We also need the 
exponentially weighted Sobolev space $H^{\{s,\delta\}}$
defined, for $s\in\{0,1,2,\dots\}$ and $\delta>0$ 
as the subspace of $L^2(\R^3)$ with the
norm
$$
\|f\|_{H^{\{s,\delta\}}}=\sum_{m:|m|=0}^s 
\|e^{\delta|x|}\partial^m f\|_{L^2}
$$
finite; we write  $H^{\{s,\delta\}}_{rad}$ for the radial part of
$H^{\{s,\delta\}}$. Since $f\in H^{\{1,\delta\}}_{rad}$ immediately 
implies $e^{\delta|x|}f\in H^1_{rad}$, with a bound for the norm, 
we have by \eqref{s4}:
\beq
\sup_{x\in\smr^3:|x|\geq R_*}|xe^{\delta|x|}f(x)|\leq
C_5\|f\|_{H^{\{1,\delta\}}}\,\qquad
\hbox{for {\em radial} functions $f$.}\la{s5}
\eeq

Introduce the Banach spaces
\begin{align}
X&=\R\times
BC^{1,2}_2\times BC^{1,1}\times 
\bigl(H^{\{2,\delta\}}_{rad}\cap BC^{1,0}\bigr)\times 
\bigl(H^{\{1,\delta\}}_{rad}\cap BC^{1,0}_1\bigr)\la{dx}
\\
Y&=BC^2_2\times BC^1\times 
\bigl(H^{\{1,\delta\}}_{rad}\cap BC\bigr)\times 
\bigl(H^{\{0,\delta\}}_{rad}\cap BC\bigr),\la{dy}
\end{align}
where for an intersection of two normed
spaces we use the norm $\|f\|_{A\cap B}=\|f\|_{A}+\|f\|_{B}$. 
Now notice that \eqref{er1}-\eqref{dr2} define 
a system of equations
${\mathcal F}(l,Q,N,\Phi_1,\psi_2;\epsilon)=0$, where 
$\mathcal{F}:X\times\R\to Y$ maps
$(l,Q,N,\Phi_1,\psi_2)\in X$ and $\epsilon\in\R$ (assumed small) to
\begin{align}
\Biggl(
(rQ)'+2l (rf_0)'-8\pi Gr^2\rho_\epsilon,\;\;&
2rN'-\frac{2l f_0+Q}{1-\epsilon^2 (2l f_0+Q)}-
\frac{8\pi G\epsilon^2 r^2 p_\epsilon}
{{1-\epsilon^2 (2l f_0+Q)}},\notag\\
&\frac{2m}{\hbar}\psi_2
+\frac{\partial\Phi_1}{\partial r}
-F_2,\;\;
\bigl(\frac{\eta}{\hbar}-\frac{mN}{\hbar}\bigr)\Phi_1
-\frac{\partial\psi_2}{\partial r}-
\frac{2}{r}\psi_2-F_1
\Biggr)\notag
\end{align}
in $Y$. 
\begin{remark}
As promised in remark \ref{r2} the field $Q$ is $O(r^{-2})$ at infinity
(since we are solving for $Q\in BC^{1,2}_2\subset BC^2$),
so that the parameter $l$ encoding the ADM mass is well defined,
by integration of the first equation:
$$
2l=\int_0^\infty 8\pi G r^2\rho_\epsilon dr.
$$
\end{remark}
\begin{remark}\la{vq}
Notice that $Q\in BC^{1,2}_2$, so that $Q=O(r^2)$ as $r\to 0$. To see
that this is natural, integrate the first equation up to $r$, to
deduce $rQ+2l rf_0(r)=\int_0^r 8\pi G s^2\rho_\epsilon ds=O(r^3)$,
and recall that we chose $f_0$ to be $O(r^2)$ as $r\to 0$.
\end{remark}
\begin{remark}
\la{wd}
The fact that ${\cal F}(\,\cdot\,;\,\epsilon)$ 
maps $X$ into $Y$ (for 
small $\epsilon>0$) can be read off
from its definition and the formulae \eqref{f1}-\eqref{reps},
by means of the inequalities \eqref{s1}-\eqref{s5} above.
For example consider the third component of ${\cal F}$:
$\psi_2$ and $\frac{\partial\Phi_1}{\partial r}$
are clearly in $H^{\{1,\delta\}}_{rad}\cap BC$, so it remains to consider
$F_2$. Referring to \eqref{f2}, it follows from the fact that
both $Q$ and $f_0(r)$ are $O(r^2)$, 
that the first term is continuous at $r=0$,
while the same is obviously true for the second and third terms,
since $\psi_2\in BC^{1,0}_1$ and $N\in BC^{1,1}$.
Furthermore,
functions
in $H^{\{1,\delta\}}$ are continuous away from the origin
and exponentially decreasing by \eqref{s5}, so that
altogether $F_2\in BC$. Similarly $F_2\in 
H^{\{1,\delta\}}_{rad}$: the only term for
which this is not immediately evident is $\frac{1}{r}\Phi_1$, but this
is premultiplied by $Q+2lf_0=O(r^2)$ as $r\to 0$. These observations
show that ${\cal F}$ is continuous from 
$X\times\R\to Y$ (locally, near $\epsilon=0$); 
this will be strengthened to locally $C^1$ in lemma \ref{smap}.
\end{remark}

\subsection{ \normalsize The Newtonian limit and statement of the main theorem} When $\epsilon=0$ 
the formal Newtonian limit solution mentioned 
previously corresponds to the observation that
\beq
\mathcal{F}\bigl(2mG,2ru'-4mGf_0 ,u  ,\varphi  , \frac{-\hbar}{2m}\varphi' 
;0\bigr)=0
\eeq
where $\varphi$ and $u$ solve \eqref{nss}. To see this, observe that
by \eqref{reps} and the preceding remarks, the equation
${\mathcal F}(l,Q,N,\Phi_1,\psi_2;0)=0$ amounts to
\begin{align}
&(rQ)'+2l (rf_0)'-16m\pi Gr^2|\Phi_1|^2=0,\label{ern1}\\
&2rN'-(2l f_0+Q)=0,\label{ern2}\\
&\bigl(\frac{\eta}{\hbar}-\frac{mN}{\hbar}\bigr)\Phi_1
-\frac{\partial\psi_2}{\partial r}-
\frac{2}{r}\psi_2=0,\label{drn1}\\
&\frac{2m}{\hbar}\psi_2
+\frac{\partial\Phi_1}{\partial r}
=0.\label{drn2}
\end{align}
The first two equations imply $-\Delta N=-8\pi mG|\Phi_1|^2$, and
substituting this into \eqref{drn1} and then substituting for $\psi_2$
from \eqref{drn2},
implies that $(\Phi_1,N)$ solves \eqref{nss}. 
We choose, for $\epsilon=0$,
$(\Phi_1,N)$to be the energy minimizing solution $(\varphi,u)$ 
of \eqref{nss}
described in section \ref{nes}. The multipole expansion
for $N=u$ and the requirement that $Q\in BC^{1,2}$ imply that
$l=2mG$, since $\int|\varphi(x)|^2 dx=1$. Collecting this together
we write $\Xi_N=\bigl(2mG,2ru'-4mGf_0 ,u  ,\varphi  , 
-\frac{\hbar}{2m}\varphi'\bigr)$
and call this the Newtonian limit point in $X$. We can now state
the main theorem:
\begin{theorem}\la{main}
There exists an interval $(-\epsilon_1,\,+\epsilon_1)$ on which
is defined a $C^1$ curve 
$\epsilon\to \Xi^\epsilon
=(l^\epsilon,Q^\epsilon,N^\epsilon,\Phi_1^\epsilon,\psi_2^\epsilon)
\in X$ of solutions to the Einstein-Dirac system \eqref{er1}-\eqref{dr2}
such that $\|\Xi^\epsilon-\Xi_N\|_X=O(\epsilon)$. More explicitly,
$l^\epsilon\to 2mG$ and
\begin{align}\la{ps1}
\|Q^\epsilon+4mGf_0-2ru'\|_{BC^{1,2}_2}+\|N^\epsilon-u\|_{BC^{1,1}}
+\|\Phi^\epsilon_1&-\varphi\|_{H^{\{2,\delta\}}_{rad}\cap BC^{1,0}}\\
&+\|\psi_2^\epsilon
+\frac{\hbar}{2m}\varphi'\|_{H^{\{1,\delta\}}_{rad}\cap BC^{1,0}_1}
=O(\epsilon).\notag
\end{align}
\end{theorem}
In terms of the original variables of the problem, we 
define a metric 
$$
g^\epsilon=\epsilon^{-2}e^{2\nu^\epsilon}dt^2
-e^{2\lambda^\epsilon}dr^2-r^2 d\Omega^2
$$ 
where
$\nu^\epsilon=\epsilon^2 N^\epsilon$ and
$(1-e^{-2\lambda^\epsilon})=\epsilon^2({2l^\epsilon}f_0(r)+Q^\epsilon)$. 
Define also $\Phi_2^\epsilon=\epsilon\psi_2^\epsilon$,
then for $\epsilon=c^{-1}$ small 
we have a solution $(\lambda^\epsilon,\nu^\epsilon,\Phi_1^\epsilon,
\Phi_2^\epsilon)$ to \eqref{d1}-\eqref{e2}, and
\begin{align}\la{ps2}
\epsilon^{-2}\|(1-e^{-2\lambda^\epsilon})&-2\epsilon^2ru'\|_{BC^{1,2}_2}
+\epsilon^{-2}\|\nu^\epsilon-\epsilon^2 u\|_{BC^{1,1}}\\
&+\|\Phi^\epsilon_1-\varphi\|_{H^{\{2,\delta\}}_{rad}\cap BC^{1,0}}
+\epsilon^{-1}\|\Phi_2^\epsilon
+\epsilon\frac{\hbar}{2m}\varphi'\|_{H^{\{1,\delta\}}_{rad}\cap BC^{1,0}_1}
=O(\epsilon).\notag
\end{align}
\begin{remark}
We briefly discuss regularity of the solutions just obtained, 
omitting the $\epsilon$ index for clarity.
The definition of the
last two factors in $X$ ensures that $\Phi_1,\psi_2$
are $C^1$ for $r>0$, and $\psi_2=O(r)$ near $r=0$; notice that
\eqref{dr2} implies that 
$\frac{\partial\Phi_1}{\partial r}=O(r)$ as
$r\to 0$ also. It follows that the corresponding 
Dirac spinor given 
by \eqref{anz2} is also $C^1$ on $\R^3$. To see this, 
first oberve that the partial derivatives at $r=0$ of the 
function $x^i\psi_2/r$ exist and given by
$\partial_j(x^i\psi_2)(0)=\delta_{ij}\psi_2'(0)$. But also
calculate
$$
\partial_j\Bigl(\frac{x^i\psi_2}{r}\Bigr)=\frac{\delta_{ij}}{r}\psi_2
+\frac{x^ix^j}{r^2}\Bigl(\psi_2'-\frac{\psi_2}{r}\Bigr),
$$
and notice that this is continuous everywhere, 
including at $r=0$ since
$$\psi_2'(0)=\lim_{r\to 0}\frac{\psi_2(r)}{r}
=\lim_{r\to 0}\psi_2'(r).$$
Also the functions $e^{2\lambda}$ and $N$ are $C^1$ functions for $r>0$,
whose derivatives have limit zero as $r\to 0$, and hence define $C^1$
functions on $\R^3$. Thus overall we have a classical $C^1$ solution
of the equations of motion.
\end{remark}
\subsection{ \normalsize Linearization and proof of main theorem} Our aim is to
use the implicit function theorem to obtain solutions of ${\cal F}=0$ 
for small $\epsilon$,
thus proving theorem \ref{main}. 
To achieve this it is sufficient to check that ${\cal F}$ is
locally $C^1$ and
its partial derivative in the $X$ direction, $D_1\mathcal{F}$,
evaluated at the Newtonian limit is a bounded linear bijection
(and hence a linear homeomorphism by the open mapping theorem).
Let $B_1(\Xi)$ be the ball of unit radius centered at $\Xi$ 
in the Banach space $X$.
\begin{lemma}
\la{smap}
For $\epsilon_*$ sufficiently small the map
$\mathcal{F}$
is $C^1$ in the neighbourhood 
$B_1(\Xi_N)\times(-\epsilon_*,\epsilon_*)$
of the Newtonian limit point
$(\Xi_N;0)$.
The partial derivative
$D_1\mathcal{F}(\Xi_N ;0)$
is the linear map $X\to Y$ which takes 
$\xi=(\delta l,q,n,\chi_1,\chi_2)\in X$ to
\begin{align}
\Biggl((rq)'+2\delta l (rf_0)'
-32\pi mGr^2\langle\varphi,&\chi_1\rangle,\;
2rn'-{2\delta l f_0}-q,\notag\\
&
\frac{2m}{\hbar}\chi_2
+\frac{\partial\chi_1}{\partial r},\;\;
\bigl(\frac{\eta}{\hbar}-\frac{mu}{\hbar}\bigr)\chi_1
-\frac{\partial\chi_2}{\partial r}-
\frac{2}{r}\chi_2-\frac{mn}{\hbar}\varphi
\Biggr)
\notag\end{align}
in $Y$. 
\end{lemma}
\proof
This can mostly be read off by inspection. Consider
for example the first component:
$(rQ)'+2l (rf_0)'-8\pi Gr^2\rho_\epsilon$. This will
define a smooth map from $X\times\R$ to $BC^2$ if each of the
three terms does so separately.
Clearly
$Q\mapsto (rQ)'$ is a continuous linear map
$BC^{1,2}_2\to BC^{2}$, while
the function $(rf_0(r))'$ lies in $BC^2$, so that 
$l\mapsto l (rf_0)'$ is continuous and linear from $\Real$ to
$BC^2$. Next, from the 
formula \eqref{reps}
it is clear (using the inequalities \eqref{s1}-\eqref{s5}) that
$\rho_\epsilon\in BC^2$, and that the corresponding map
is $C^1$ (in fact smooth) from $X\times\R$ to $BC^2$. 
The second component of $\mathcal{F}$
is handled similarly, after choosing $\epsilon_*$
to be sufficiently small that 
$|{{1-\epsilon^2 (2l f_0+Q)}}|>\frac{1}{2}$ on 
$B_1(\Xi_N)\times(-\epsilon_*,\epsilon_*)$.
The third and fourth components can also be handled
in a straightforward way, noting that the coefficients
in the expressions \eqref{f1}-\eqref{f2} 
(i.e. the quantities multiplying $\Phi_1,\psi_2$
and their derivatives) are smooth
functions on $B_1(\Xi_N)\times(-\epsilon_*,\epsilon_*)$ 
taking their values in $BC^{1,1}$; see also remark \ref{wd}.
\myqed
\begin{lemma}
Let $\varphi$ be an energy minimizing solution of \eqref{nss} as
described in theorem \ref{lt}, and let $\Xi_N$ be the corresponding
Newtonian limit point satisfying $\mathcal{F}(\Xi_N;0)=0$ as
just described.
Then the partial derivative in the $X$ direction,
$D_1\mathcal{F}(\Xi_N;0)$,
is a linear homeomorphism 
$X\to Y$.
\end{lemma}
\proof 
To establish this we need to prove the 
unique solvability (with bounds for $\xi\in X$)
of the equation $D_1\mathcal{F}(\Xi_N;0)(\xi)=y$,
for $y=(\alpha,\beta,j_1,j_2)\in Y$. Thus we consider the system
\begin{align}
(rq)'+2\delta l (rf_0)'-32\pi mGr^2\langle\varphi,\chi_1\rangle&=\alpha,\la{q}\\
2rn'-{2\delta l f_0}-q&=\beta,\la{n}\\
\frac{2m}{\hbar}\chi_2
+\frac{\partial\chi_1}{\partial r}&=j_1,\la{c2}\\
\bigl(\frac{\eta}{\hbar}-\frac{mu}{\hbar}\bigr)\chi_1
-\frac{\partial\chi_2}{\partial r}-
\frac{2}{r}\chi_2-\frac{mn}{\hbar}\varphi&=j_2.
\la{c1}
\end{align}
To start with, multiply the second equation by $r$,
differentiate, add the first equation and then divide by $2r^2$, 
leading to
\beq\la{poi}
-\Delta n+16\pi mG\langle\varphi,\chi_1\rangle
+\frac{\alpha+(r\beta)'}{2r^2}=0.
\eeq
Now use the representation \eqref{fs} for $n$, 
and substitute the
resulting formula for $n$ into the equation obtained by
eliminating $\chi_2$ between the third
and fourth equations:
$$
-\frac{\hbar^2}{2m}\Delta\chi_1-\eta\chi_1+mu\chi_1+mn\varphi
=-\hbar j_2-\frac{\hbar^2}{2m}\Bigl(
\frac{\partial}{\partial r}+\frac{2}{r}\Bigr)j_1.
$$
This yields
$$
L\chi_1=J
$$
where $L$ is the operator in lemma \ref{lll} and
$J$ is the function
$$
J(r)=-\hbar j_2
-\frac{\hbar^2}{2m}\Bigl(\frac{\partial j_1}{\partial r}
+\frac{2j_1}{r}\Bigr)
+m\varphi\biggl[(-\Delta)^{-1}\bigl(\frac{\alpha+(r\beta)'}{2r^2}\bigr)
\biggr].
$$
In order to apply lemma \ref{lll} we need to check
that $J\in L^2_{rad}$:
the first two terms clearly are, so we just need
to show the same is true of the third term.
Let $\dot H^1_{rad}$ be the homogeneous Sobolev space 
with norm $\|f\|_{\dot H^1}^2=\int|\nabla f|^2 d^3x$, and let
$\dot H^{-1}_{rad}$ be the dual space. Recall that 
$\dot H^1_{rad}\subset L^6_{rad}$
with a continuous embedding, by Sobolev's inequality.
Now $\alpha/r\in L^2_{rad}$, so 
$r^{-2}\alpha\in \dot H^{-1}_{rad}$ since,
by Hardy's inequality,
$$
|\int f \frac{\alpha}{r^2} r^2dr|\leq \|\alpha/r\|_{L^2}
\|f/r\|_{L^2}\leq C\|\alpha\|_{BC^1}
\|f\|_{\dot H^1}.
$$
(Here the $L^2$ norm is understood to be
with respect to the measure $4\pi r^2dr$ since all functions
are radial.)
Also $r^{-2}(r\beta)'\in\dot H^{-1}_{rad}$ since
$$
|\int f \frac{(r\beta)'}{r^2} r^2dr|=
|\int {r\beta}f' dr|\leq 
\|\beta/r\|_{L^2}
\|f'\|_{L^2}\leq C\|\beta\|_{BC^1}
\|f\|_{\dot H^1}.
$$
Given this it follows from the Riesz representation theorem that
$(-\Delta)^{-1}\bigl(\frac{\alpha+(r\beta)'}{2r^2}\bigr)\in\dot H^1_{rad}
\subset L^6_{rad}$ is well-defined, and the bound
$$
\bigl\|(-\Delta)^{-1}\bigl(\frac{\alpha+(r\beta)'}{2r^2}\bigr)\bigr\|
_{\dot H^1\cap L^6}\leq C\|(\alpha,\beta)\|_{BC^1\times BC^1}
$$
holds. Therefore $\varphi(-\Delta)^{-1}
\bigl(\frac{\alpha+(r\beta)'}{2r^2}
\bigr)\in L^2_{rad}$ 
(since $\varphi$ is smooth and decays exponentially
and so is in $L^3_{rad}$, and $\frac{1}{2}=\frac{1}{3}+\frac{1}{6}$).
As a consequence there 
exists a unique
$\chi_1\in H^2_{rad}$ such that $
L\chi_1=J
$
satisfying the bound 
$$
\|\chi_1\|_{H^2}\leq C\Bigl(\|(\alpha,\beta)\|_{BC^1\times BC^1}
+\|j_2\|_{L^2}+\|j_1\|_{H^1}\Bigr).
$$
To derive the exponentially weighted bound, we apply
lemma \ref{exph2} below to the equation:
$$
(-\Delta + M + V(r))\chi_1=f=\frac{2m}{\hbar^2}\Bigl[
-\hbar j_2-\frac{\hbar^2}{2m}\bigl(j_1'+\frac{2}{r}j_1\bigr)
-mn\varphi\Bigr]
$$
where $M=-2m\eta/\hbar^2>0$ and $V=2m^2u/\hbar^2$. This
implies
\begin{align}
\|\chi_1\|_{H^{\{2,\delta\}}}&\leq c\Bigl(\|\chi_1\|_{H^2}+
\|f\|_{H^{\{0,\delta\}}}\Bigr)\notag\\
&\leq
c\Bigl(
\|(\alpha,\beta)\|_{BC^1\times BC^1}
+\|j_2\|_{H^{\{0,\delta\}}}+\|j_1\|_{H^{\{1,\delta\}}}
+\|n\|_{L^6}\|e^{\delta r}\varphi\|_{L^3}
\Bigr)\notag
\end{align}
for $\delta$ small ($\delta<\sqrt{M}$ and less than the decay rate of
$\varphi$ in theorem \ref{lt}).
The third equation above then determines 
$\chi_2\in H^{\{1,\delta\}}_{rad}$ uniquely:
$
\chi_2
=\frac{\hbar}{2m}(j_1
-\frac{\partial\chi_1}{\partial r})$, with 
the same bound for $\|\chi_2\|_{H^{\{1,\delta\}}}$
as for $\|\chi_1\|_{H^{\{2,\delta\}}}$.

We show that also $\chi_2\in BC_1^{1,0}$ below. First we must 
discuss $q$ and $n$.
Since we are attempting to solve for $q=O(r^{-2})$ as
$r\to +\infty$, integration of the first equation gives
$$
2\delta l=\int_0^\infty
\bigl(32\pi mGr^2\langle\varphi,\chi_1\rangle+\alpha\bigr)dr
$$
and
\begin{align}
q(r)&=2\delta l\bigl(\frac{1}{r}- f_0(r)\bigr)-\frac{1}{r}\int_r^\infty
\bigl(32\pi mGs^2\langle\varphi(s),\chi_1(s)\rangle+\alpha(s)\bigr)ds
\notag\\
&=-\frac{1}{r}\int_r^\infty\Biggl[
-2\delta l \bigl(sf_0(s)\bigr)'+
\bigl(32\pi mGs^2\langle\varphi(s),\chi_1(s)\rangle+\alpha(s)\bigr)
\Biggr]ds
\notag\\
&=+\frac{1}{r}\int_0^r\,\,\Biggl[
-2\delta l \bigl(sf_0(s)\bigr)'+  
\bigl(32\pi mGs^2\langle\varphi(s),\chi_1(s)\rangle+\alpha(s)\bigr)
\Biggr]ds\notag
\end{align}
Since $\chi_1$ is bounded and continuous, the first line makes
it clear that $q=O(r^{-2})$ as $r\to +\infty$, while the third
implies that $\lim_{r\to 0}r^{-2}q(r)$ exists (since
$\alpha\in BC^2_2$), and hence that $q\in BC^{1,2}_2$. 
Recalling that $\varphi$ is
exponentially decaying, and
putting together the obvious
estimates gives the bounds 
$
|\delta l|\leq 
c(\|\alpha\|_{BC^2}+\|\chi_1\|_{H^2})
$ 
and
$
\|q\|_{BC^2_2}\leq c(|\delta l|+\|\alpha\|_{BC^2_2}+\|\chi_1\|_{H^2}).
$
The similar bound for $\|q'\|_{BC^3_1}$ then follows from 
\eqref{q}, yielding the bound for
$\|q\|_{1,2;2}$ the norm on $BC^{1,2}_2$. 
Next for $n$ which was defined to solve \eqref{poi};
we must show it satisfies \eqref{n}. So 
let $H=2rn'-{2\delta l f_0}-q-\beta$, 
then \eqref{poi} and \eqref{q} imply that $(rH)'=0$
so that $H=c/r$. But $r^{-1}H\in L^{2}_{rad}$ 
(since $n'\in L^2_{rad}$)
so $c=0$, and hence $H=0$ i.e. \eqref{n} holds. This
implies immediately that $n'\in BC^2$ and $n\in BC^1$
with a bound $\|n\|_{BC^{1,1}}\leq c(|\delta l|+\|\beta\|_{BC^1}
+\|q\|_{BC^1}).$

Next define $\chi_2\in H^{\{1,\delta\}}_{rad}$ by \eqref{c2};
it follows that \eqref{c1} holds, and so $\chi_2$ is $C^1$
for $r>0$.
To establish $\chi_2\in BC_1^{1,0}$ it is necessary
to analyze the behaviour at the origin. To achieve this,
integrate up the fourth equation:
$$
\chi_2(r)=\frac{1}{r^2}\int_0^{r}
\biggl[
\bigl(\frac{\eta}{\hbar}-\frac{mu}{\hbar}\bigr)\chi_1
-j_2-\frac{mn}{\hbar}\varphi\biggr]s^2ds.
$$
The quantity in square brackets is continuous by assumption,
so that $\lim_{r\to 0}r^{-1}\chi_2(r)$ exists and is finite; it
follows from \eqref{c1} that $\lim_{r\to 0}\chi'_2(r)$ exists
and is finite also, so that $\chi_2\in BC^{1,0}_1$ as required.
Finally \eqref{c2} then implies that $\chi_1\in BC^{1,0}$.\myqed

\begin{lemma}[Exponentially weighted bounds]\la{exph2}
Assume $u\in H^2_{rad}(\R^3)$ solves
$$
\bigl(-\Delta+M+V(r)\bigr)u=f
$$
where $f\in H^{\{0,\delta\}}_{rad}(\R^3)$, 
for some $\delta<\sqrt{M}$. Assume further that
$V$ is continuous 
and $\lim_{r\to+\infty}V(r)=0$. Then
$u$ also satisfies the bound $\|u\|_{ H^{\{2,\delta\}}}\leq
c\bigl(\|u\|_{H^2}+\|f\|_{H^{\{0,\delta\}}}\bigr)$
for some $c=c(\delta)>0$.
\end{lemma}
\proof
$v=ru$ solves
\beq\la{dg}
-v''+Mv+Vv=rf 
\eeq
For any $S>R+1>R$ let $b(r)$ be a 
smooth function with $b(r)=0$ if $r\leq R$ or $r\geq S+1$, 
and $b(r)=1$ if
$S>r\geq R+1$. Multiply by $e^{2\delta r}b(r)$ and integrate, 
estimate
$
\bigl|\int 2\delta e^{2\delta r}bvv' dr\bigr|\leq
\int \bigl(e^{2\delta r}\delta^2bv^2/(1-\epsilon)
+(1-\epsilon)e^{2\delta r}bv'^2\bigr)dr
$
and integrate by parts all other terms involving $vv'$. 
This leads to
$$
\int_0^\infty\Bigl[
\epsilon e^{2\delta r}b v'^2
+(M-\frac{\delta^2}{(1-\epsilon)}+V)b e^{2\delta r} v^2
\Bigr]dr
\leq c\int_0^\infty\Bigl[
\delta e^{2\delta r}|b'| v^2+e^{2\delta r} |b''|{v^2}
+re^{2\delta r}|bfv|\Bigr]dr.
$$
For any $\epsilon>0$ let $R$ be such that $\sup_{r\geq R} |V(r)|
<\epsilon$, and let 
$\delta^2<{(M-2\epsilon)(1-\epsilon)}$, so that
$|M-\frac{\delta^2}{(1-\epsilon)}+V|>\epsilon$ for $r\geq R$.
Now let $S\to+\infty$, to deduce
\beq
\int_{R+1}^\infty\Bigl[
e^{2\delta r}v'^2+e^{2\delta r} v^2
\Bigr]dr\leq c(\epsilon)\bigl(\|u\|^2_{L^2}+
\int_R^\infty
re^{2\delta r}|fv|dr
\bigr).
\la{st}\eeq
Using the fact that $\|u\|^2_{ H^{\{1,\delta\}}}\leq
c\bigl(\|u\|^2_{H^1}+\int_{R+1}^\infty e^{2\delta r}(v^2+v'^2)dr\bigr)$,
\eqref{st} implies $\|u\|_{ H^{\{1,\delta\}}}\leq
c\bigl(\|u\|_{H^1}+\|f\|_{H^{\{0,\delta\}}}\bigr)$. To improve this
to $H^{\{2,\delta\}}$ it is only necessary to multiply the equation
\eqref{dg} by $e^{{\delta}{r}}b^{\frac{1}{2}}$, 
square and integrate, to
obtain 
$$
\int_{R+1}^\infty e^{2\delta r}v''^2dr
\leq c\Bigl(\|u\|^2_{ H^{\{0,\delta\}}}
+\|f\|^2_{ H^{\{0,\delta\}}}\Bigr)
$$
since $\|f\|^2_{ H^{\{0,\delta\}}}=4\pi\int r^2e^{2\delta r}f^2 dr.$
But $\|u\|^2_{ H^{\{2,\delta\}}}\leq
c\bigl(\|u\|^2_{ H^{\{1,\delta\}}}+\int_{R+1}^\infty 
e^{2\delta r} v''^2dr\bigr)$
since for an arbitrary second order derivative
$r^2|\nabla_i\nabla_j u|^2\leq cr^2(|u''|^2+r^{-2}|u'|^2)$
so that in the region $r>R+1>1$ there holds
$r^2|\nabla_i\nabla_j u|^2\leq c(|v''|^2+r^2|u|^2+r^2|u'|^2)$.
This completes the proof.
\myqed



\end{document}